# How Do Proteins Fold?

Carlos Bustamante[1,2,3,4,5,6], Christian Kaiser[7,8], Erik Lindahl[9,10], Robert Sosa[1], Giovanni Volpe[11,12]

1. Department of Molecular and Cell Biology, University of California, Berkeley, Berkeley, CA 94720, USA
2. California Institute for Quantitative Biosciences, University of California, Berkeley, Berkeley, CA 94720, USA
3. Jason L. Choy Laboratory of Single-Molecule Biophysics, University of California, Berkeley, Berkeley, CA, USA
4. Department of Physics, University of California, Berkeley, Berkeley, CA, USA
5. Howard Hughes Medical Institute, University of California, Berkeley, Berkeley, CA, USA
6. Kavli Energy Nanoscience Institute, University of California, Berkeley, Berkeley, CA, USA
7. Department of Biology, Johns Hopkins University, Baltimore, MD, USA
8. Bijvoet Centre for Biomolecular Research, Department of Chemistry, Utrecht University, Utrecht, The Netherlands
9. Department of Applied Physics, Science for Life Laboratory, KTH Royal Institute of Technology, Stockholm, Sweden
10. Department of Biochemistry and Biophysics, Science for Life Laboratory, Stockholm University, Stockholm, Sweden
11. Physics Department, University of Gothenburg, 412 96 Gothenburg, Sweden
12. SciLifeLab, Physics Department, University of Gothenburg, 412 96 Gothenburg, Sweden

**Abstract**

How proteins fold remains a central unsolved problem in biology. While the idea of a folding code embedded in the amino acid sequence was introduced more than 6 decades ago, this code remains undefined. While we now have powerful predictive tools to predict the final native structure of proteins, we still lack a predictive framework for how sequences dictate folding pathways. Two main conceptual models dominate as explanations of folding mechanism: the *funnel model*, in which folding proceeds through many alternative routes on a rugged, hyperdimensional energy landscape; and the *foldon model*, which proposes a hierarchical sequence of discrete intermediates. Recent advances on two fronts are now enabling folding studies in unprecedented ways. Powerful experimental approaches—in particular, single-molecule force spectroscopy and hydrogen–deuterium exchange assays—allow time-resolved tracking of the folding process at high resolution. At the same time, computational breakthroughs culminating in algorithms such as AlphaFold have revolutionized static structure prediction, opening opportunities to extend machine learning toward dynamics. Together, these developments mark a turning point: for the first time, we are positioned to resolve how proteins fold, why they misfold, and how this knowledge can be harnessed for biology and medicine.

**Main Text**

In the late 1950s, seminal experimental studies by Christian Anfinsen led him to propose the "thermodynamic hypothesis" of protein folding, stating that the native state of a polypeptide corresponds to its global free-energy minimum under a given set of external conditions (Anfinsen *et al.*, 1961; Epstein *et al.*, 1963). The subsequent observation that many proteins fold *in vitro* without any external aid by simply being placed in close-to-physiological conditions implied that all the information required to determine their three-dimensional folded native structures is encoded in the protein's sequence (Anfinsen, 1973). This realization led to a great effort among biophysicists and biochemists to solve the "protein folding problem", which can be formulated as a set of three related questions (Dill *et al.*, 2008): (1) How is folding encoded by the amino acid sequence? (2) What is the folding mechanism? (3) Is it possible to predict tertiary structures (including those of intermediates) from a protein's amino acid sequence?

The studies that followed Anfinsen's pioneering work used site-directed mutagenesis to systematically determine how and where the information needed to attain the native state is encoded (Matthews, 1996). They suggested that the contributions of individual amino acid residues to the attainment of the folded state depend on the *sequence context* where those residues are found. Certain residues were found to be crucial for the stability and, in some cases, the foldability of the polypeptide chain, whereas others appeared to have little influence on folding overall, merely modulating folding kinetics or thermodynamic stability (Shortle, 1989; Baase *et al.*, 2010). Clustering and packing of hydrophobic side chains in the interior of globular proteins was recognized as a main contribution for stabilizing folded structures and guiding folding (Kauzmann, 1959; Nozaki and Tanford, 1971; Dill, 1985). Yet, these hydrophobic interactions are necessary but not sufficient for polypeptides to fold into native proteins, and almost seven decades of experimental and theoretical inquiry have not revealed a "folding code" at the amino acid level, i.e., rules endowed with the generality and predictive power required to connect amino acid sequence to how the protein attains its structure.

The failure to describe how proteins fold is perhaps better illustrated when compared to our current understanding of the folding of ribonucleic acids (RNAs) (Tinoco and Bustamante, 1999). Simple yet powerful rules of engagement resulting from base pairing of purines and pyrimidines lead to the formation of secondary structures such as helices, bulges, internal and hairpin loops, which then arrange in space to arrive at a tertiary structure. This process makes it possible to describe the energy function or Hamiltonian of the system as the additive contribution of three terms (Tinoco and Bustamante, 1999). The first, and energetically the largest, is associated with the formation of secondary structure interactions; the second, smaller in magnitude, corresponds to the contribution of attaining and stabilizing tertiary structure interactions; the third, and by far the smallest of the three, is a cross term coupling secondary and tertiary structure contributions. This energy "separability" implies that it is possible to devise a hierarchical "Aufbau" approach to arrive at the tertiary structure of RNA from its sequence (Turner & Mathews, 2010). The folding of RNA molecules can thus be conceived as a modular and hierarchical process in which folding involves transitions from primary to the secondary and finally to tertiary structures through successive processes of energy minimization.

In the case of proteins, the rules of secondary structure prediction from the interactions of amino acid residues are significantly weaker and more complex than in nucleic acids. Moreover, the stability of secondary structures (alpha helices and beta sheets) mainly involves the backbone and not directly the side chains. The side chains play a role in the tertiary packaging and their complexity is great. In the 1960s, Lifson introduced the concept of theoretically trying to calculate the probability of secondary structures (Lifson, 1961), but it was not until the 1970s that Chou and Fasman were able to empirically derive an alpha helix prediction algorithm by statistical analysis of amino acid propensities in experimentally determined structures (Chou and Fasman, 1974). The predictions for beta sheets or for random coiled regions proved more difficult. This challenge is due in part to the fact that alpha helices are mainly stabilized by local interactions between adjacent residues, while sheets depend on non-local interactions.

So far, it has not been possible to design a hierarchical scheme to go from the primary to secondary structure of a protein, and from there to its tertiary structure. Several reasons contribute to the difficulty of developing an analytical model of the relationship between sequence and folding, including the enormous multiplicity of possible interactions among the 20 amino acid building blocks, the lack of an energy database associated with these interactions both within and between secondary structures, and the fact that the stability of a given secondary structure in a protein depends strongly on the tertiary context where it is found in the native structure. This latter circumstance implies that the third or coupling term of the Hamiltonian, which in the RNA case is the smallest of the three contributions, is not negligible for proteins, and as a result the separability of the Hamiltonian is no longer possible. Since around the turn of the century, it has even been possible to simulate the entire process for some small proteins from physical interaction models (Duan and Kollman, 1998) to, e.g., understand the role of water in folding (Rhee *et al.*, 2004), but at very high computational cost, and the proteins targeted have been selected or even engineered to be small and extremely fast-folding, rather than representative of actual proteins found in nature.

While a folding code has not yet been defined, structure prediction algorithms have continuously improved (Baker 2019), and a giant step in our ability to predict 3D structure entirely from sequence was recently accomplished with the development of AlphaFold (Jumper *et al.*, 2021; Abramson *et al.*, 2024). A key factor enabling this success has been to exploit the large databases of experimentally determined protein sequences and structures (Berman *et al.*, 2000; Benson *et al.*, 2000). Machine learning made it possible to identify weak correlations to generate the structure most likely to correspond to a sequence. This tour-de-force effort has largely solved the problem of predicting protein structure from sequence (the third question in the protein folding problem), but with a key limitation: the algorithm that predicts the structure is a complex black box of pattern recognition that casts little light on the process of folding and that tells us nothing about why only some sequences fold, or how physics and evolution are coupled. Indeed, the folding code is clearly reflected in evolutionary conservation and co-variances, but we lack the ability to read it. As a result, crucial aspects of protein folding still remain in the dark: Why do specific sequence changes, including disease-causing mutations, result in misfolding? Where along the folding trajectory does the process go awry? What properties of naturally occurring proteins necessitate help from molecular chaperones to reach folding efficiencies that are satisfactory to sustain cellular fitness? And how may proteins assume several distinct structures, either intrinsically or upon interacting with partner molecules?

Answering these questions requires knowledge not only of the final structure, but also of the pathway leading to it, spurring a search for models that explain the folding process. Thus, despite the recent breakthroughs described above, structure prediction alone does not resolve the central question of how folding occurs. Anfinsen's dictum—namely that all the information required by a protein to attain its native state is encoded in its sequence—remains true. But how does a folding protein find its native structure among the astronomically large number of possible conformations? The complexity of this search problem was captured in Levinthal's paradox (Levinthal, 1968), which pointed out that a random search over all accessible conformations would take longer than the age of the universe whereas, in reality, most proteins fold within milliseconds to seconds. This apparent contradiction implies that folding cannot be a random search but must instead proceed along defined pathways. So, how does a protein find its native state without being marred in Levinthal's paradox?

The *energy landscape model* (Leopold *et al.*, 1992; Bryngelson *et al.*, 1995; Dill and Chan, 1997) offers a convenient solution to Levinthal's paradox. This model views the conformational search as a diffusion over a funnel-shaped energy surface. The term "funnel" was introduced by Ken A. Dill in 1987 (Dill, 1987). The energy landscape relates the microscopic degrees of freedom of the folding protein to its free energy. In the funnel representation of the potential energy surface, as shown in Figure 1A, the depth corresponds to the stabilization free energy of the molecule relative to the unfolded state at the top of the funnel, and the width represents the multiplicity of conformational states of the system. In their search for the native state, which resides in the narrow bottom of the funnel, molecules diffuse over the topographic details of the energy surface, driven by minimization of the free energy. In principle, an infinite number of possible folding pathways on this surface are possible for the molecule to reach its native state, as long as those paths move in a direction that decreases the free energy of the molecule, bringing it ever closer to the free energy minimum where the native state resides. Local minima in the topography of the funnel surface account explicitly for the formation of productive folding intermediates and kinetically trapped off-pathway states. The funnel model provides an intuitive solution to Levinthal's paradox by suggesting the inevitability of the folding of any protein that diffuses on the surface defined by the funnel and by providing a huge number of alternative pathways to do so.

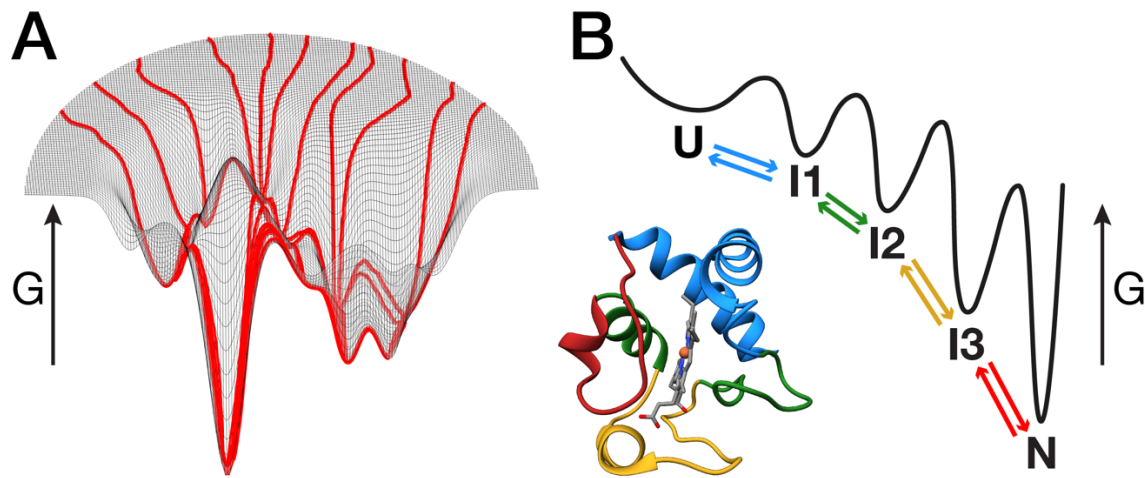

**Figure 1 | Conceptual models of protein folding: funnel versus foldon.** Schematic comparison of the two dominant frameworks for protein folding. (**A**) *Funnel model*: proteins diffuse over a rugged energy landscape toward the native state at the bottom of the funnel, with multiple possible pathways and potential kinetic traps. Folding information is treated as being distributed along the sequence, allowing initiation at many points near the rim of the funnel. (**B**) *Foldon model*: proteins fold hierarchically through discrete units of a few tens of residues, termed "foldons", that adopt structure in a strict order to generate a well-defined folding pathway. The two models embody contrasting views (robustness through multiple pathways versus reliability through ordered intermediates) and distinguishing between them is a central challenge in the field.

Indeed, the funnel model implies that a protein can start folding by forming stabilizing contacts in any region of its sequence. The unit of folding information in this model is the amino acid residue, and any combination of these residues can function as a nucleation site to initiate folding, as long as the process decreases the free energy of the system and brings it closer to the native state at the bottom of the funnel. In this way, the picture of the funnel model implies that the information needed to fold is distributed more or less evenly throughout the protein sequence. No group of amino acids has a defined priority over any other group. The protein, thus, can initiate folding anywhere, and it proceeds to organize its residues in a manner that continuously moves the system towards the bottom of the funnel. The funnel model implies a true random search for the native state in the accessible configurational space of the protein, guided by free energy minimization. It can be thought of as a diffusion with an energetic drift towards the bottom of the funnel.

Despite its attractive, intuitive description of the folding process, several considerations argue against the "distributed" model implied by the funnel model. For example, the reduction in free energy as the molecule diffuses on the funnel-shaped energy landscape does not necessarily imply a concomitant increase in the number of native contacts of the protein. Formation of intermediate structures with low free energy but a small fraction of native contacts is perfectly possible in this model. Because of their stability (low free energy), structures can represent significant kinetic traps and impediments for the attainment of the native state. In other words, the multiplicity of pathways implicit in the funnel model predicts a large degree of frustration (Ferreiro *et al.*, 2018)—many if not most of the pathways would lead to kinetically trapped states, which will make the attainment of the native state a highly unlikely event.

It is known that proteins that attain well-defined folded structures can be subjected to rather major "surgery" and still retain their ability to reach their native state. One example of such procedures is circular permutation, in which the N- and C-termini of the protein are joined together and the protein new N- and C-termini are generated by splitting the polypeptide somewhere else along the chain. The circular permutant's primary structure shows a dramatic change in amino acid sequence and yet, in many instances, the polypeptide retains the ability to attain the original native state. How can this observation be reconciled with the idea that the sequence of the chain determines the native state?

An alternative view, supported by experimental observations of a limited set of model proteins (Bai *et al.*, 1995; Englander and Mayne, 2014; Englander *et al.*, 2016), posits the existence of defined folding pathways. According to this model, proteins do not start folding at any arbitrary place along the sequence, but do so in discrete steps that involve groups of a few tens of residues (typically about 20 to 30 residues). These groups, termed "foldons" by Englander and collaborators, represent the actual folding units (Figure 1B). The existence of foldons was initially formulated to rationalize observations from hydrogen–deuterium exchange (HDX) studies of cytochrome c folding (Bai *et al.*, 1995). In this type of experiment (Figure 2A), folding of a deuterated protein is initiated in conditions in which the proteins can exchange amide-deuterons for protons. The reaction is stopped at different times, and the protein is then digested and analyzed by mass spectrometry (Figure 2A). Regions of the protein that fold first are protected from exchange more than those that fold later. In these experiments, cytochrome c formed always the same intermediates and always in the same order of appearance, defining a unique folding pathway. For cytochrome c to have a well-defined, unique folding pathway, it is necessary that its constituent folding units have a defined folding order; i.e., its foldons must possess a folding *hierarchy* in the protein. Similar conclusions were reached based on HDX studies of another protein, ribonuclease H (Hu *et al*. 2013) (Figures 2B and 2C), suggesting that this mode of folding might be generally applicable.

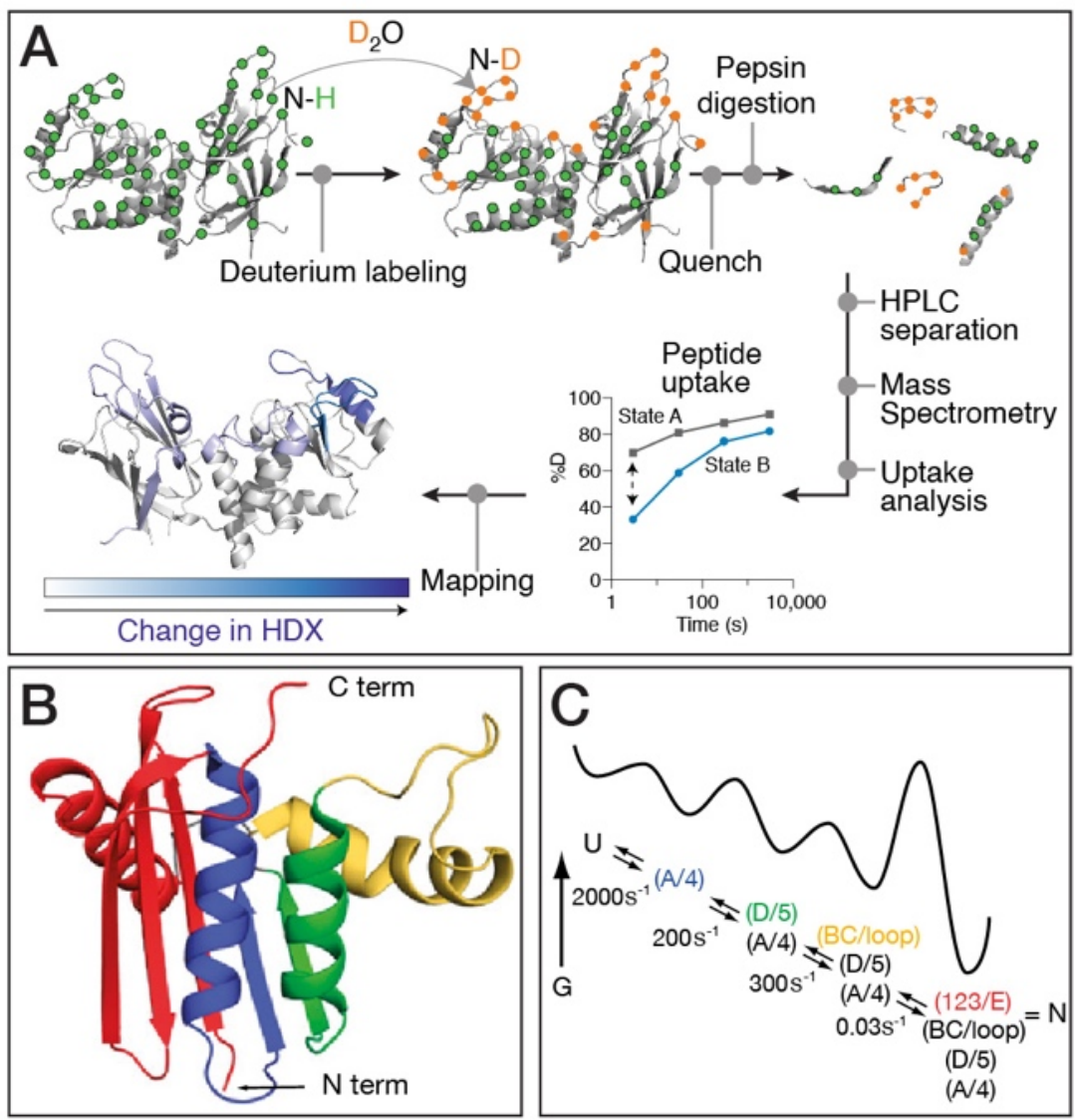

**Figure 2 | Hydrogen–deuterium exchange experiments to map folding pathways.** (**A**) HDX workflow. Deuterium is incorporated into the protein backbone during labeling, before exchange is quenched by a shift to acidic pH. The protein is then proteolytically cleaved and analyzed by mass spectrometry. Segments that were not stably structured during labeling show increased exchange. (**B**) Time-resolved HDX measurements of RNaseH folding reveal individual foldons (color-coded as on panel C) that structure sequentially. (**C**). Schematic one-dimensional energy diagram of RNase H folding, based on experimental data. Modified from Masson *et al.* 2019 (panel A) and Hu *et al.* 2013 (panels B, C).

The *foldon model* offers a plausible explanation to some of the observations that are difficult to rationalize otherwise, starting from why it has not been possible to find simple folding rules of which amino acids in a protein play a major role in defining folding pathways and which do not. Experiments of single amino acid mutagenesis cannot find those rules if the folding units are not the individual residues but a set of them as postulated by the foldon model. The folding model also explains why products of circular permutation often arrive at the same native state with similar stability. If the joining of the original N- and C-termini and subsequent generation of a new set of termini do not affect the ability of the protein's foldons to adopt a secondary or tertiary structure, the protein could still move through the same folding intermediates to arrive at the folded state.

What the foldon model implies is that the idea that the amino acid sequence determines the tertiary structure of proteins as postulated by Anfinsen must be refined. A folding protein, according to this model, must be seen not as a "polymer of amino acids", but as an "oligomer of its constituent foldons". Note also that the foldon concept brings back, at least in part, the modular or hierarchical folding we associate with RNA described above. The foldon model proposes that proteins fold by first organizing themselves into small secondary and partial tertiary structures whose sequences can be thought to encode the necessary interactions to adopt those structures. Because the folding of these units must occur in a strict hierarchy, there must be *primary* foldons whose tendency to adopt their folded structure ensures that the protein starts folding there. The primary foldons are followed by secondary foldons whose structural organization depends on and requires not only their own intra-foldon interactions but also interactions with foldons of higher hierarchy, which fold before them. The result is a process that leads to a well-defined path of ordered folding intermediates.

From an evolutionary perspective, the idea that certain short sequences emerged early by virtue of possessing a strong tendency to adopt secondary or tertiary structures and that increasingly more complex forms arose from a combination of them, appears as an attractive scenario for how proteins eventually replaced the relatively simpler folding RNA structures. The existence of foldons also explains why the nearly 250,000 structures currently deposited in the Protein Data Bank (Berman *et al.*, 2000; Burley *et al.*, 2025) consistently contain a small number of motifs (e.g., helix-turn-helix, helix-turn-beta, beta-alpha-beta, beta hairpin, beta barrels) that repeat in various combinations over and over again. It seems reasonable to assume that their existence reflects the fact that protein polymers have a natural tendency to fold in small units that often adopt the same spatial organization despite having different amino acid sequences.

It is tempting to imagine the emergence of folding in evolution as a hierarchical phenomenon, starting with the appearance of independent folding units that form globular structures and continuing in similar fashion at higher scales of complexity. In fact, protein architecture seems to be organized in a modular manner: as proteins become larger, they tend to segregate into globular domains. This segregation reduces folding complexity and ensures minimal frustration (Dill, 1985). These segregated domains have often evolved to perform specialized functions as seen in allosteric proteins and, often, they attain their native form in a well-defined order relative to each other, requiring some degree of inter-domain coupling (Sun *et al.*, 2011; Shank *et al.*, 2009). It seems therefore reasonable to think of these domains as larger cooperative folding elements, a sort of "super-foldons".

At the time of writing, the controversy remains unresolved: Is protein folding a multiple pathway process whose robustness arises precisely from the many alternative ways to attain the native state as postulated by the funnel model? Would such a mechanism, on the contrary, promote trapping and frustration? Or has the folding of proteins evolved, instead, along single paths punctuated by the attainment of well-defined intermediates built up from the successive addition of smaller folding units, always arising in the same order until the native state is reached as postulated by the foldon model? Understanding how proteins fold is not just a problem of academic interest, it would also make it possible to establish the mechanisms that lead to misfolding, a problem of great medical importance (Dobson, 2001; Valastyan & Lindquist, 2014; Sontag *et al.*, 2017).

What is needed is a way to discriminate between these folding models. We need to be able to distinguish whether every time the molecule folds it follows a different pathway or if it always describes the same path, punctuated by its obligatory attainment of the same discrete intermediates. The hydrogen–deuterium exchange method has been applied to a discrete set of proteins and has so far supported the channeled or foldon model (Bai *et al.*, 1995; Hu *et al.*, 2013). However, this is an ensemble approach and because of the difficulty to synchronize a population of molecules as they undergo a dynamic process, it is unclear whether the observed intermediates always correspond to discrete and distinct thermodynamic states or if they represent a redistribution of the unfolded ensemble when exposed to conditions that favor folding.

Single molecule force spectroscopy represents an interesting alternative (Bustamante and Yan, 2022)**.** The capability of this method to follow in real time the folding trajectory of a single protein has made it possible to observe distinct protein folding intermediates (Figure 3). Because these exist transiently, they are not easy to characterize. However, a method that would permit the annotation of these intermediates would represent an important way to determine if these intermediates are always the same, if they are on path to the native state, and if they appear in the same order every time the protein folds, as suggested by the foldon model. This is the challenge.

On the computational side, significant challenges remain in bridging the gap between static structure prediction and dynamic folding pathways. All-atom molecular dynamics can, in principle, capture folding trajectories, but the timescales and system sizes of biologically relevant proteins make such simulations prohibitively expensive even on modern supercomputers (Abraham et al., 2015). Coarse-grained and enhanced-sampling methods, such as Markov state models or replica-exchange techniques, can extend accessible timescales but at the cost of reduced resolution (Husic & Pande, 2018). More recently, deep learning approaches inspired by the success of AlphaFold offer new opportunities to learn the statistical patterns of folding dynamics from evolutionary, structural, and experimental data (Lewis et al., 2025). Integrating these computational strategies with single-molecule experiments, high-throughput assays, and hydrogen–deuterium exchange will be essential to map out folding landscapes at sufficient resolution and scale. Ultimately, progress will depend on developing hybrid frameworks in which simulations and machine learning are tightly coupled to experimental benchmarks, allowing us to distinguish between competing models and uncover general principles of the mechanism of protein folding.

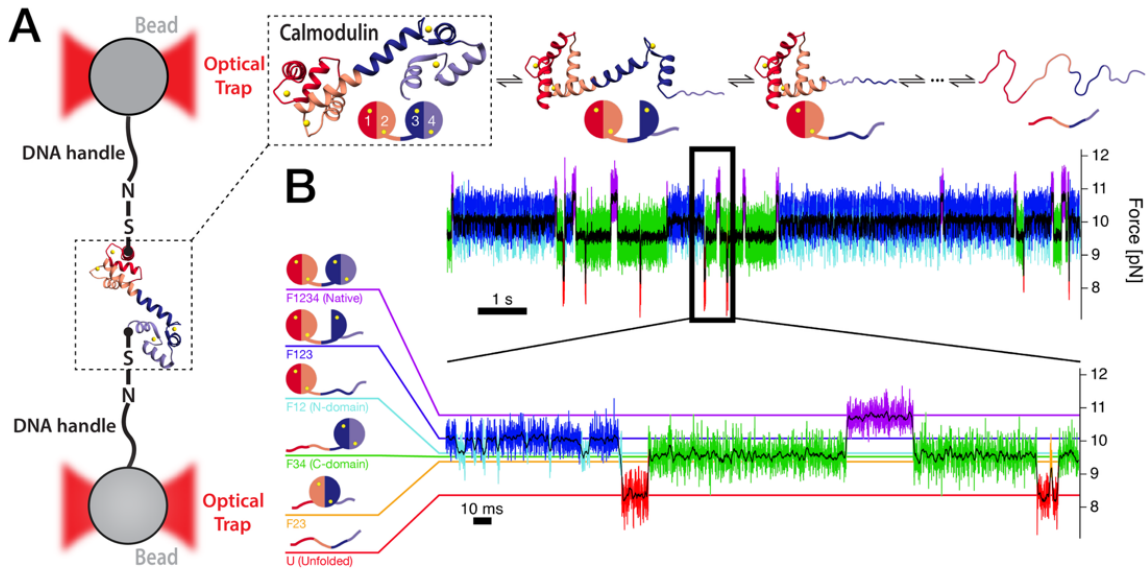

**Figure 3 | Single-molecule analysis of protein folding.** (**A**) Experimental setup. A single protein molecule (calmodulin, composed of four EF hand motifs) is tethered between two beads held in optical traps that serve as force probes. (**B**) Holding the molecule at an appropriate tension, structural transitions in the molecule are reflected transitions between discrete forces. The high temporal resolution enables observation of transitions between states in which the EF hands are unfolded (U), fully structured (F1234), partially folded (F123, F12, F34) or misfolded (F23), kinetically resolving the complex folding network of calmodulin. Modified from Stigler *et al.* 2011.

Looking ahead, the resolution of the protein folding problem will come from the synergy of experimental and computational advances. Single-molecule tools (Bustamante *et al.*, 2020; Petrosyan *et al.*, 2021; Zhang *et al.*, 2025), high-throughput cellular assays (Goldman *et al.*, 2015; Chen *et al.*, 2025), and hydrogen–deuterium exchange (Englander & Mayne, 2014; Masson *et al.*, 2019) now allow us to observe folding intermediates directly, while machine learning and large-scale simulations are beginning to capture the complexity of folding landscapes (Lewis et al., 2025). The next frontier is to combine these approaches: experiments will generate rich, high-resolution data that can train and validate models, while computational frameworks will provide hypotheses and predictive power that guide experimental design. This iterative loop between measurement and prediction promises not only to distinguish between competing models such as funnels and foldons, but also to reveal the principles that govern folding, misfolding, and the emergence of functional diversity. By uniting physics-based insight with data-driven learning, we envision a future where protein folding is no longer a paradox, but a solvable, predictive science with transformative implications for biology, medicine, and design.

**References**

Abraham, M. J., T. Murtola, R. Schulz, S. Páll, J. C. Smith, B. Hess & E. Lindahl (2015). "GROMACS: High performance molecular simulations through multi-level parallelism from laptops to supercomputers." *SoftwareX* 1:19–25.

Abramson, J., J. Adler, J. Dunger, R. Evans, T. Green, A. Pritzel, O. Ronneberger, L. Willmore, A. J. Ballard, J. Bambrick & S. W. Bodenstein (2024). "Accurate structure prediction of biomolecular interactions with AlphaFold 3." *Nature* 630(8016): 493–500.

Anfinsen, C. B., Haber, E., Sela, M., & White, F. H. (1961). "The kinetics of formation of native ribonuclease during oxidation of the reduced polypeptide chain." *PNAS* 47(9): 1309–1314.

Anfinsen, C. B. (1973). "Principles that govern the folding of protein chains." *Science* 181(4096): 223–230.

Baase, W. A., L. Liu, D. E. Tronrud & B. W. Matthews (2010). "Lessons from the lysozyme of phage T4." *Protein Science* 19(4): 631–641.

Bai, Y., T. R. Sosnick, L. Mayne & S. W. Englander (1995). "Protein folding intermediates: native-state hydrogen exchange." *Science* 269(5221): 192–197.

Baker, D. (2019). "What has de novo protein design taught us about protein folding and biophysics?" *Protein Science* 28(4): 678–683.

Baldwin, R. L. (2012). "The protein folding funnel: A primer with experiments." *Proteins: Structure, Function, and Bioinformatics* 80(3): 566–570.

Benson, D. A., M. Cavanaugh, K. Clark, I. Karsch-Mizrachi, D. J. Lipman, J. Ostell & E. W. Sayers (2012). GenBank. *Nucleic Acids Research* 41(D1):D36–D42.

Berman, H.M., J. Westbrook, Z. Feng, G. Gilliland, T.N. Bhat, H. Weissig, I.N. Shindyalov & P.E. Bourne (2000). "The Protein Data Bank." *Nucleic Acids Research* 28(1):235–242.

Bryngelson, J. D., J. N. Onuchic, N. D. Socci & P. G. Wolynes (1995). "Funnels, pathways, and the energy landscape of protein folding: a synthesis." *Proteins* 21(3): 167–195.

Burley, S. K., R. Bhatt, C. Bhikadiya, C. Bi, A. Biester, P. Biswas, S. Bittrich, S. Blaumann, R. Brown, H. Chao, V. Reddy Chithari, P. A. Craig, G. V. Crichlow, J. M. Duarte, S. Dutta, Z. Feng, J. W. Flatt, S. Ghosh, D. S. Goodsell, R. Kramer Green, V. Guranovic, J. Henry, B. P. Hudson, M. Joy, J. T. Kaelber, I. Khokhriakov,J.-S. Lai, C. L. Lawson, Y. Liang, D. Myers-Turnbull, E. Peisach, I. Persikova, D. W. Piehl, A. Pingale, Y. Rose, J. Sagendorf, A. Sali, J. Segura, M. Sekharan, C. Shao, J. Smith, M. Trumbull, B. Vallat, M. Voigt, B. Webb, S. Whetstone, A. Wu-Wu, T. Xing, J. Y. Young, A. Zalevsky & C. Zardecki (2025). "Updated resources for exploring experimentally-determined PDB structures and Computed Structure Models at the RCSB Protein Data Bank." *Nucleic Acids Research 53*(D1):D564–D574.

Bustamante, C., L. Alexander, K. Maciuba & C. M. Kaiser (2020). "Single-molecule studies of protein folding with optical tweezers." *Annual Review of Biochemistry* 89(1):443–470.
10